\documentclass[article,10pt,onecolumn,superscriptaddress,floatfix]{revtex4}
\usepackage{amsmath,amssymb,graphicx,units}
\usepackage[usenames,dvipsnames]{color}
\usepackage{lipsum}
\usepackage{setspace}
\usepackage{mathrsfs}
\usepackage[bookmarks=false,linkcolor=blue,urlcolor=blue,colorlinks,citecolor=blue]{hyperref}

\begin{document}
\title{Magnetization jump in one dimensional $J-Q_{2}$ model with anisotropic exchange}
\author{Bin-Bin Mao}
\affiliation{Center of Interdisciplinary Studies $\&$ Key Laboratory for Magnetism and Magnetic Materials of the Ministry of Education, Lanzhou
University, Lanzhou 730000, China}
\author{Chen Cheng}
\affiliation{Beijing Computational Science Research Center, Beijing 100193, China}
\author{Fu-Zhou Chen}
\affiliation{Center of Interdisciplinary Studies $\&$ Key Laboratory for Magnetism and Magnetic Materials of the Ministry of Education, Lanzhou
University, Lanzhou 730000, China}
\author{Hong-Gang Luo}
\email{luohg@lzu.edu.cn}
\affiliation{Center of Interdisciplinary Studies $\&$ Key Laboratory for Magnetism and Magnetic Materials of the Ministry of Education, Lanzhou
University, Lanzhou 730000, China}
\affiliation{Beijing Computational Science Research Center, Beijing 100193, China}
\begin{abstract}
We investigate the adiabatic magnetization process of the one-dimensional $J-Q_{2}$ model with XXZ anisotropy $g$ in an external magnetic field $h$ by using density matrix renormalization group (DMRG) method. According to the characteristic of the magnetization curves, we draw a magnetization phase diagram consisting of four phases. For a fixed nonzero pair coupling $Q$, i) when $g<-1$, the ground state is always ferromagnetic in spite of $h$; ii) when $g>-1$ but still small, the whole magnetization curve is continuous and smooth; iii) if further increasing $g$, there is a macroscopic magnetization jump from partially- to fully-polarized state; iv) for a sufficiently large $g$, the magnetization jump is from non- to fully-polarized state.
By examining the energy per magnon and the correlation function, we find that the origin of the magnetization jump is the condensation of magnons and the formation of magnetic domains. We also demonstrate that while the experienced states are Heisenberg-like without long-range order, all the \textit{jumped-over} states have antiferromagnetic or N\'eel long-range orders, or their mixing.
\end{abstract}
\maketitle
\section{Introduction}
Quantum spin systems play a very important role in condensed matter physics, because of their underlying rich physics, such as the spin liquid state~\cite{Anderson1987} and the valence-bond solid (VBS) state~\cite{Read1989}. Typically, subjected to external magnetic field, the magnetization process of the spin systems can exhibit anomalous phenomena. Among them two kinds of nonanalytic magnetization behaviors have attracted many interests. One is the magnetization plateau, which usually accompanies with the spin excitation gap and has been found in many systems, such as the frustrated spin systems~\cite{Ono2003,Honecker2004}, and quasi-periodic systems with nontrivial topological property~\cite{Hida1993, Hu2014}. The other is the magnetization jump, which exhibits discontinuity in the magnetization density.

The magnetization jump was first proposed by N\'eel~\cite{ap_neel} in the system with the Ising-like anisotropic exchange interaction, and then also investigated in various of lattice spin systems in different dimensions ~\cite{Kohno1997,Sakai1999,Aligia2000,Schulenburg2002,Dmitriev2006,Kalinov2006,Heidrich-Meisner2009,Kolezhuk2012,Albarracin2014,Kishine2014,Hiroki2015,Morita2016}. Most of these model systems involve anisotropy or frustration. Experimentally, the magnetization jump was first confirmed in hydrated copper compound ~\cite{Poulis1951}, and then was found in many kinds of magnetic materials ~\cite{Moller1977,Hardy2003,Ghivelder2004,Yoshii2007,Diop2016,Manago2016,Maji2010}. However, understanding the mechanism of the magnetization jump in an intuitive way is still in exploration. Many explanations have been presented for this issue, such as the magnetic domain reorientation~\cite{Moller1977,Hardy2003,Maji2010}, the spin-flop transition~\cite{Gerhardt1998, Sakai1999,Hiroki2015}, the formation of bound magnon pairs~\cite{Dmitriev2006}, and the macroscopically large degeneracy at the critical value of the external magnetic field~\cite{Schulenburg2002,Richter2004}.

Recently, field driven phase transition has been proposed in one-dimensional (1D) $J-Q_{2}$ model~\cite{Adam2015,Adam2016}. This model was first introduced by Sandvik~\cite{Sandvik2007} to construct a spin valence-bond-solid (VBS) state without frustration.
In the presence of external magnetic field, the numerical results by employing the exact diagonalization and the stochastic series expansion quantum Monte Carlo (QMC) method~\cite{Syljuasen2002} show that the magnetization curve of the model displays a sharp jump from a finite value to the saturated magnetization density at certain critical magnetic field.
In their work, the origin of the magnetization jump is explained as the onset of attractive interactions between magnons, according to the analytical results for two magnons on a ferromagnetic background. However, one notes that the anisotropic exchange effect, which is usually closely related to the magnetization jump, has not been considered in Ref.~\cite{Adam2016}.
In the present work, we numerically investigate the one-dimensional $J-Q_{2}$ model with XXZ anisotropy using DMRG method. We obtain a novel anisotropy dependent magnetization phase diagram with considerable physics.
It shows that the magnetization jump behavior can be evidently influenced (either depressed or enhanced) by anisotropy.
Interestingly, if the anisotropy strength $g$ is large enough, e.g. $g>4$ in units of $J$, a direct jump from a non-polarized to a fully-polarized state occurs.
We emphasize that this direct magnetization jump observed in the strongly anisotropic case is found for the first time, which is absent in the isotropic one.
We systemically explore the mechanism of the magnetization jump in the whole parameter regime by analysing the properties of $N$-magnon state, i.e. its ground state energy, correlation function and long-range order. Focusing on the excitation energy per magnon for $N$-magnon state and the corresponding excitation energy difference between (\textit{N}+1)- and \textit{N}-magnon states, we determine the critical magnetization density and external magnetic field at which the magnetization jump appears.
Analysis of system's energy in the whole magnetization process indicates that magnetic domain forms in the \textit{jumped-over} states. This reveals that the magnetization jump shown in this work is due to the formation of the magnetic domain, in which region all spins are in a uniform direction. 
This understanding is also supported by analytical calculations in some limit cases, e.g., $g\rightarrow\infty$ and few magnon limit. 
In addition to energetic consideration, we further analyse the correlation function for each magnetization sector and different  parameters. 
We find that while the experienced states in the magnetization process are Heisenberg-like without long-range order, all the \textit{jumped-over} states have antiferromagnetic or N\'eel long-range orders, or their mixing.

The paper is organized as follows. In the following section we introduce the anisotropic $J-Q_{2}$ model and the numerical method we used. In the section ``Results", the magnetization jump behavior in different parameter regimes is illustrated and a novel anisotropy induced phase diagram is presented.
In the section ``Discussion", we analyse the mechanism of the magnetization jump both in the few magnon limit and the whole magnetization process.

\section{Model Hamiltonian and Numerical Method}
\label{sec_2}
The anisotropic $J-Q_{2}$ model in the presence of an external magnetic field is described by the Hamiltonian
\begin{equation}\label{ham_jq}
H=-J\sum_{i}P_{i,i+1}-Q\sum_{i}P_{i,i+1}P_{i+2,i+3}-h\sum_{i}S_{i}^{z},
\end{equation}
where $P_{i,j}\equiv\frac{1}{4}-\left(S_{i}^{x}S_{j}^{x}
+S_{i}^{y}S_{j}^{y}+g S_{i}^{z}S_{j}^{z}\right)$ and $g$ is the $XXZ$ anisotropy. $J$ is the Heisenberg exchange constant, $Q$ is the coupling strength of the nearest pairs, and $h$ is the external magnetic field. $g =1$ recovers the isotropic limit.
In the isotropic limit without magnetic field, the competition between $J$ and $Q$ terms leads to a ground state phase transition from Heisenberg ground state to the doubly degenerate VBS phase~\cite{Tang2011}. In this paper, we are more interested in the adiabatic magnetization process of the system subjected to the external magnetic field.
To describe the magnetization process, we define the magnetization density as
\begin{equation}
m=\frac{2}{L}\sum_{i}\langle S_{i}^{z}\rangle,
\end{equation}
where $L$ is the system size. It can be readily obtained that it is always fully magnetized ($m = 1$) if $g < -1$. On the other hand, $g > -1$ is a non-trivial case, in which we can analytically stress the behavior of $m$ in some limit cases: $m=0$ for $h=0$ and $m=1$ if $h$ is large enough.
In this work, we explore how the magnetization density $m$ extrapolates from zero to saturation between these two limits. Of course, calculating $m$ in a general value of $h$ should resort numerical ways.

In practice, we numerically employ the density matrix renormalization group (DMRG) method {\cite{white1992, white1993}}, which is extremely powerful for the one-dimensional systems. We perform the calculation for systems with different lattice sizes up to 240, to obtain the physics in the thermodynamic limit. The periodic boundary condition (PBC) is adopted and the DMRG many-body states $M$ are kept dynamically \cite{Legeza2003} in order to control the truncation error. In DMRG calculations, the computational cost is in the order of $M^3$. There are two ways to choose $M$, one is to fix $M$, in this case the truncation error is different for different steps. The other is to fix truncation error, and in this case the number of the kept many-body states changes. In this work, in order to reduce the computational cost, we choose the latter, and dynamically control $M$ up to 2000, to guarantee the truncation error $\varepsilon<10^{-8}$ in the whole calculations we performed. 
In the rest of the paper, we use $J=1$ as the energy scale and restrict $Q$ and $h$ to positive values.

\section{Results}
\label{sec_3}

\begin{figure}[!tb] 
	\centering
	\includegraphics[width=0.9\columnwidth]{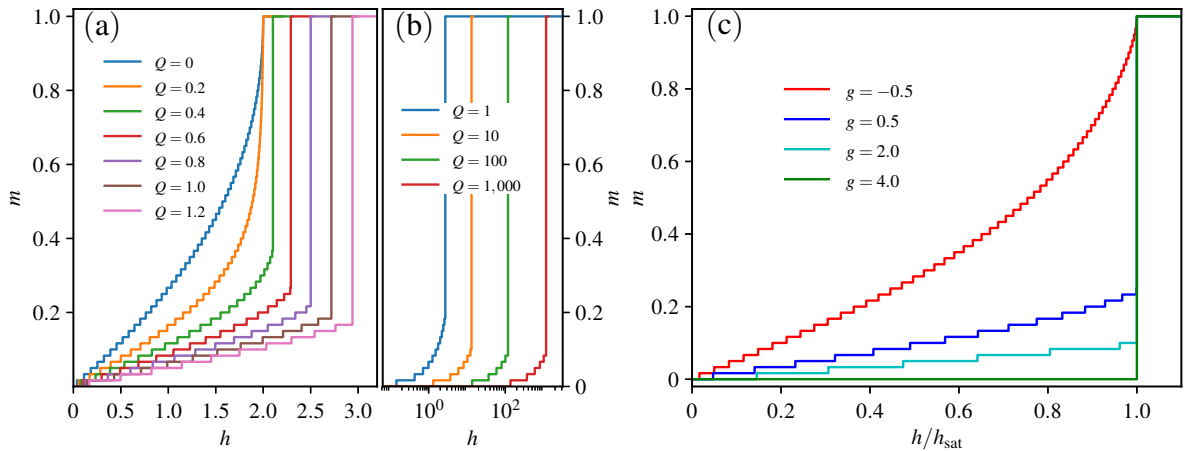}
	\caption{DMRG calculation of the magnetization density $m$ as a function of external field $h$. The system size $L=120$. (a-b) The isotropic case with $g=1$ and different coupling $Q$. (c) The anisotropic case exampled by $Q=1.5$ in different $g$. Here $h_{\rm sat}$ is the critical field when the magnetization density $m$ goes to its saturated value. }
	\label{fig:mz}
\end{figure}

\begin{figure}[!tb] 
	\centering
	\includegraphics[width=0.9\columnwidth]{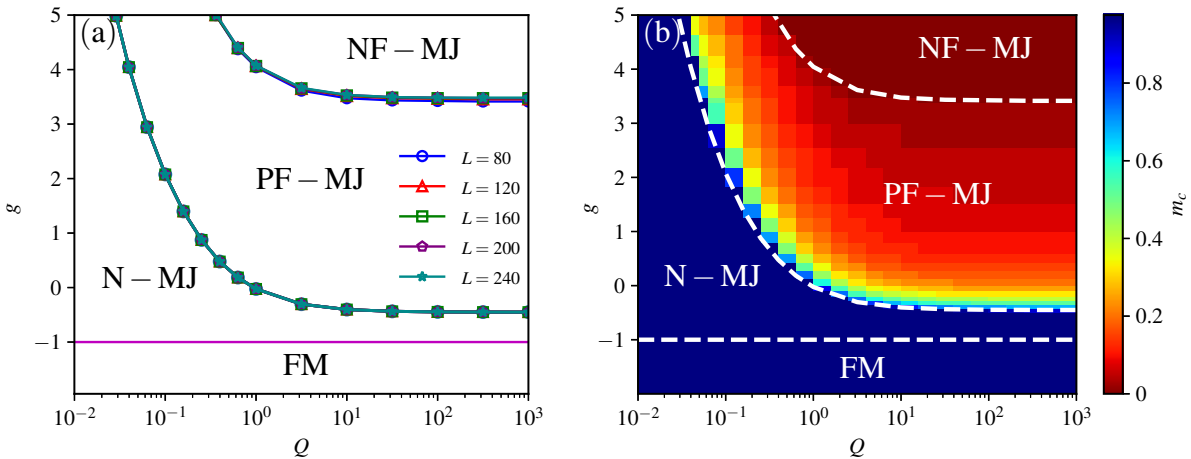}
	\caption{ Magnetization phase diagram consisting of four phases according to the behaviors of the magnetization jump processes: i) the ferromagnetic (FM) phase; ii) the no magnetization jump (N-MJ) phase; iii) the partially- to fully-polarized magnetization jump (PF-MJ) phase; iv) the non- to fully-polarized magnetization jump (NF-MJ) phase. (a) Magnetization phase boundaries with different system size. (b) Magnetization phase diagram shown by the critical magnetization $m_c$ in \{$Q,g$\} space with system size $L=80$. The white dashed-lines are phase boundaries for $L=80$ for comparison.}
	\label{fig:phase_diag}
\end{figure}

We first revisit the magnetization property in the isotropic case using DMRG calculation. The magnetization process in different strength of nearest pair coupling $Q$ is shown in Fig.~\ref{fig:mz} (a) and (b).
When $Q=0$, the system is the spin-1/2 Heisenberg chain, and its zero temperature magnetization curve is continuous and smooth. Here the small jumps and plateaus come from the finite size effect and will disappear in the thermodynamic limit. For a small $Q=0.2$, comparing to $Q=0$, $m$ changes rapidly near the saturated magnetization, but still, goes smoothly to $m=1$ at the same saturated field $h_{\rm sat}$ without a macroscopic magnetization jump. Further increasing $Q$ to $0.4$, the magnetization density $m$ changes suddenly from a partially-polarized value $m_c$ to $m=1$, and the saturated field $h_{\rm sat}$ is also larger than that for the smooth magnetization curves. This sharp jump of the magnetization curve indicates a ground state phase transition induced by the external field. The above results obtained by DMRG calculation agree with that in Ref.\cite{Adam2016} well. The difference is that in our DMRG calculation, the zero temperature case can be directly addressed, while the QMC method needs an extrapolation from finite small temperature to zero.

In the presence of a magnetization jump, the critical field $h_{\rm sat}$ increases as the coupling strength $Q$ increases, and the critical magnetization $m_c$ is smaller for a larger $Q$. However, as shown in Fig.~\ref{fig:mz}(b), even if $Q$ is sufficiently large, a finite value of $m_c$ does not decrease anymore and converges to a nonzero value. It implies no direct magnetization jump from $m=0$ to $m=1$ even when $Q\rightarrow\infty$.

Then we extend our investigation to the general case with a tunable anisotropy $g$.
Since $g < -1$ is a trivial case as mentioned above, we need just discuss the case of $g > -1$.
In Fig.~\ref{fig:mz}(c), we show the magnetization curves with a fixed typical value of $Q$ (i.e. $Q = 1.5$) in several different values of anisotropy $g$. When $g=-0.5$, the magnetization density $m$ increases gradually from $0$ to $1$ without macroscopic magnetization jump. For larger values of anisotropy $g=0.5$ and $2.0$, we can observe the shape jumps from a finite $m_c$ to the saturated magnetization density. Furthermore, when $g$ is sufficiently large ($g=4.0$), a direct jump from $m=0$ to the fully-polarized state occurs. We again point out that this novel phenomenon can not be observed in the isotropic system.

According to the different behaviors of the magnetization process, we can summarize our main results in a phase diagram consisting of four regions, as shown in Fig.~\ref{fig:phase_diag}(a). When $g<-1$, the system is in the ferromagnetic (FM) phase, and the magnetization property is trivial. When $g>-1$, the magnetization curve of the system has three different shapes: there is  i) no magnetization jump (N-MJ), ii) a partially- to fully-polarized magnetization jump (PF-MJ), iii) a non-polarized to fully-polarized magnetization jump (NF-MJ). The phase boundaries obtained by DMRG show a good convergence as the system size increases, indicating that these phases are stable in the thermodynamic limit. From these boundaries, we see that both the pair coupling $Q$ and the anisotropy $g>-1$ can enhance the magnetization jump. Furthermore, the critical anisotropy $g$ for both boundaries seems to converge in the large $Q$ limit. We show a visual variation of critical magnetization density $m_{c}$ in $\{Q,g\}$ space in Fig.~\ref{fig:phase_diag}(b), one can see that $m_{c}$ decreases with the increase of pair coupling $Q$ or anisotropy $g$ which means that the magnetization jump is enhanced.

\section{Discussion}
\label{sec_4}
\subsection{The magnetization jump in the few magnon limit}
Macroscopic magnetization jumps have been extensively discussed for various systems in the literature\cite{Heidrich2006,Kecke2007,Heidrich2009,Adam2016}. Among others, the attractive interaction between magnons plays an important role in leading to magnetization jump. For example, in the isotropic $J-Q_{2}$ model \cite{Adam2016}, Iaizzi \textit{et al.} found from QMC simulation a macroscopic magnetization jump. At the same time, their theoretical analysis for two-magnon case demonstrates that in this case these two magnons form a bound state due to an effectively attractive interaction between them. However, Iaizzi \textit{et al.} also pointed out that the formation of a bound state between two magnons is not sufficient condition for the macroscopic magnetization jump, and furthermore, an effectively attractive interaction between the magnon pairs or a cluster including macroscopic number of magnons is needed \cite{Adam2016,Heidrich2006,Kecke2007,Heidrich2009}. In order to demonstrate this, in the following we consider the few magnon limit up to four magnons.

\begin{figure}[!tb] 
	\centering
	\includegraphics[width=0.9\columnwidth]{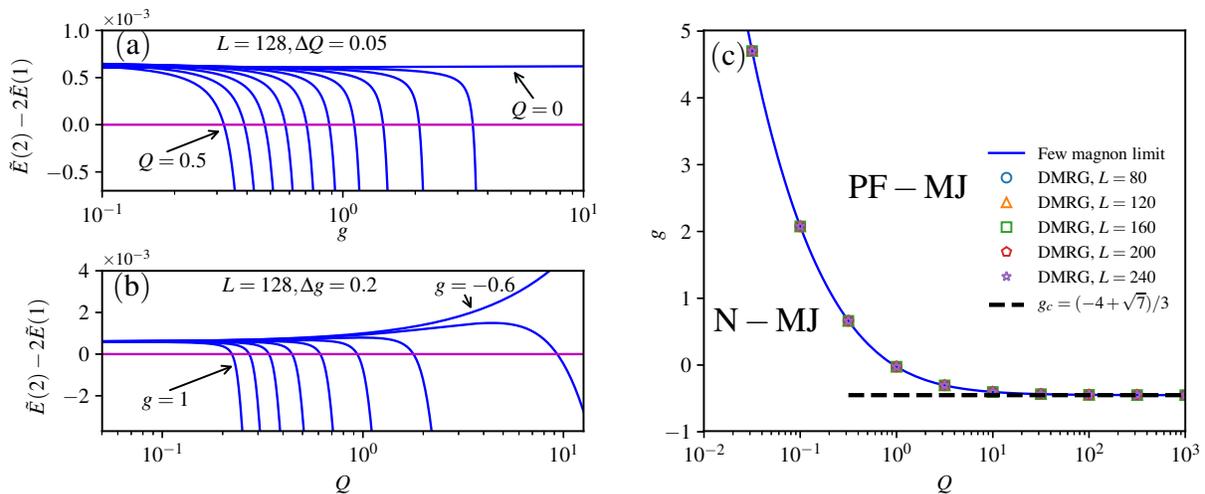}
	\caption{ $\tilde{E}(2)-2\tilde{E}(1)$  as a function of (a) $g$ for different $Q$ , (b) $Q$ for different $g$. The results are obtained by exact diagonalization in the few-magnon basis for system size $L=128$. (c) Phase boundary between the N-MJ and PF-MJ phase. The blue solid-line is obtained in the few magnon limit. The symbols are obtained using DMRG with different system sizes. The black dashed line describes the asymptotic value at which the magnetization jump appears in the large $Q$ limit.}
	\label{fig:e1e2_compare}
\end{figure}

From Fig.\ref{fig:mz} we can see that with the increase of coupling strength $Q$ or anisotropy $g$, the magnetization jump first appears near $m=1$. Thus we can analyse the origin of the magnetization jump in the ferromagnetism background. In the system with up to two magnons, we can easily get the ground state energy of the system (details in supplementary material).
For convenience, the $N$-magnon excitation energy is defined as
\begin{equation}\label{eq:tilde_E}
	\tilde{E}(N)=E(N)-E(0),
\end{equation}
where $E(N)$ is the ground state energy of the system with $N$ magnons and without external magnetic field.
The information from the value of $\tilde{E}(2)-2\tilde{E}(1)$ helps us to understand the mechanism of the magnetization jump in the few magnon limit.
The negative value of $\tilde{E}(2)-2\tilde{E}(1)$ indicates that the effective interaction between the two magnons is attractive, and thus the magnetization curve exhibits a macroscopic magnetization jump near the saturated magnetization. In contrast, if $\tilde{E}(2)-2\tilde{E}(1)>0$, the effective interaction is repulsive and there is no signal of magnetization jump for the few magnon limit. $\tilde{E}(2)-2\tilde{E}(1)=0$ is the critical case, in which the two-magnon system is in an effectively noninteracting magnon ground state.

In Fig.~\ref{fig:e1e2_compare}(a), (b), we show the results of the quantity $\tilde{E}(2)-2\tilde{E}(1)$ for the system with $L=128$, which is an example size with negligible size effect.
The magnetization density curve is smooth and continuous if the pair coupling $Q=0$ because the system has no magnetization jump according to Fig.~\ref{fig:phase_diag}.  Correspondingly, $\tilde{E}(2)-2\tilde{E}(1)$ is almost independent of $g$ and always positive as shown in Fig.\ref{fig:e1e2_compare}(a). However, for a very small $Q=0.05$, $\tilde{E}(2)-2\tilde{E}(1)$ is positive for small values of $g$, but negative when $g$ is large enough. As the anisotropy $g$ increases, the effective interaction between magnons changes from repulsive to attractive. This means the magnetization jump can be induced by the anisotropy. The boundary between the N-MJ phase and the PF-MJ phase in Fig.~\ref{fig:phase_diag} can be determined by a critical $g$ when $\tilde{E}(2)-2\tilde{E}(1)=0$. From the curves in different $Q$, we can also conclude that a needed $g$ for a magnetization jump is smaller when $Q$ is larger, in agreement with the results by DMRG (see Fig.~\ref{fig:phase_diag}).

Similar to Fig.~\ref{fig:e1e2_compare}(a), we show $\tilde{E}(2)-2\tilde{E}(1)$ as a function of $Q$ for different $g$ in Fig.~\ref{fig:e1e2_compare}(b). In the isotropic case with $g=1$, $\tilde{E}(2)-2\tilde{E}(1)>0$ for small $Q$, but becomes negative for large $Q$. A magnetization phase transition from the N-MJ phase to the PF-MJ phase occurs at the critical $Q_c(g=1)=2/9$, in agreement with the result in Ref.~\cite{Adam2016}. Notably, our large-scale DMRG calculation gives exactly the same critical $Q_c$. Different curves for decreasing $g$ show that the magnetization jump exists in the anisotropic case, and the critical value of $Q$ is larger for smaller $g$. However, when $g$ is too small ($g=-0.6$), the curve of $\tilde{E}(2)-2\tilde{E}(1)$ goes up as $Q$ increases, and there is no cross with $\tilde{E}(2)-2\tilde{E}(1)=0$. In this case, the
effective interactions between two magnons are always repulsive, and there is no signal for the magnetization jump from the two-magnon state to the saturated state. 
From Figs.~\ref{fig:e1e2_compare}(a) and (b), we can get the critical $g$ and $Q$ corresponding to $\tilde{E}_{2}-2\tilde{E}_{1}=0$. Thus we can obtain the phase boundary between N-MJ and NF-MJ phase as shown in Fig.~\ref{fig:e1e2_compare}(c).
We can see that the phase boundary obtained in the few magnon limit perfectly agrees with the numerical results by DMRG calculation.
We also notice that the asymptotic behavior of this curve can be analytically evaluated, namely, $g_c$ approaches to $\left( -4+ \sqrt{7} \right)/3$ in large $Q$ limit (details in supplementary material).

\begin{figure}[!tb] 
	\centering
	\includegraphics[width=0.9\columnwidth]{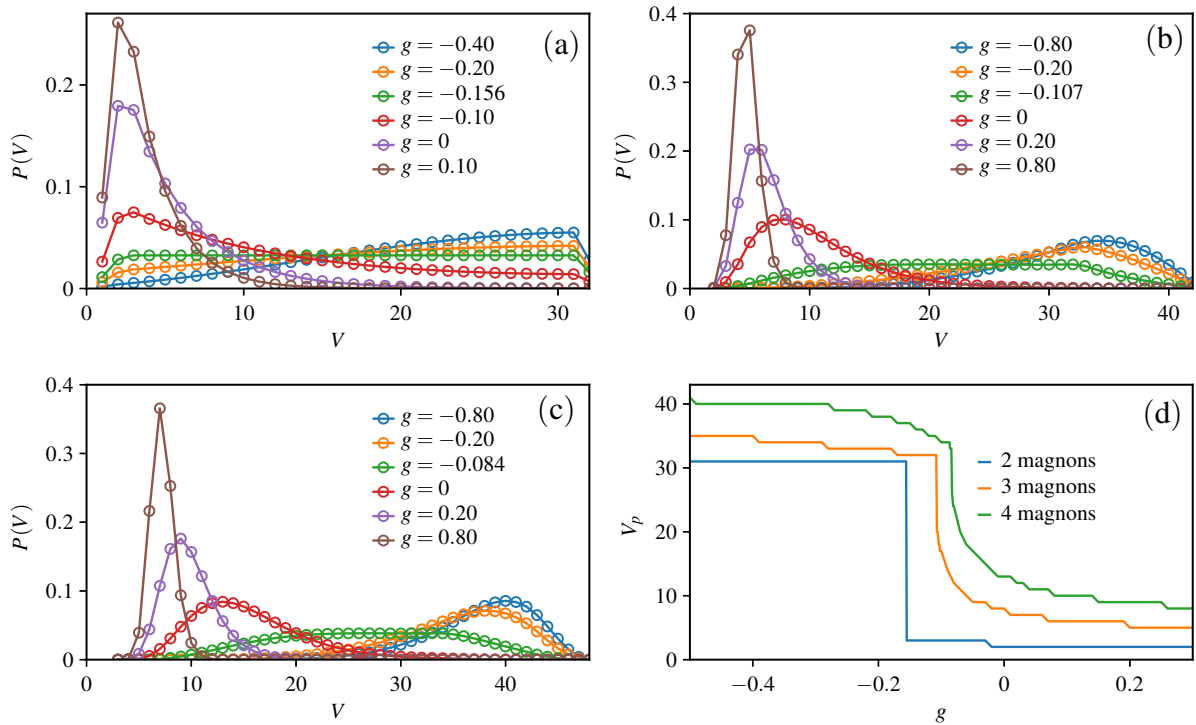}
	\caption{Probability $P(V)$ of the magnon occupied volume $V$ for the system with (a) two magnons, (b)three magnons, (c)four magnons. (d) $V_{p}$ as a function of anisotropy $g$. In the calculation we take the system size $L=64$ and $Q=1.5$.}
	\label{fig:multi_magnons}
\end{figure}

In order to further unveil the origin of macroscopic magnetization jump,  the analysis of a system with many (macroscopic number) magnons is necessary. We use $N$ intervals $d_{1},d_{2}\cdots d_{N}$, which describe the distance between the nearest neighbor magnons for an $N$-magnon state, to mark the different configurations of the $N$-magnon state. Due to the periodic boundary condition, only $N-1$ intervals are independent. Thus, to describe the distribution feature of the magnons, we define the magnon occupied volume as $V=\sum_{i}'d_{i}$\cite{Kecke2007}, where the prime means that the summation discards the largest interval. Obviously, a small value of $V$ indicates the preference of magnon condensation, while the large one corresponds to magnon separated case.

Using the exact diagonalization method, we can get the probability $P(V)$ of the system with up to four magnons.
The probability of the magnon occupied volume $V$ for a state $\left|\psi\right\rangle$ is defined as:
\begin{equation}
P(V)=\sum_{\sum_{i}^{\prime}d_{i}=V}\left|C_{d_{1},d_{2},\cdots,d_{N}}\right|^{2},
\end{equation}
where $C_{d_{1},d_{2},\cdots,d_{N}}=\left\langle \psi\big|d_{1},d_{2},\cdots,d_{N}\right\rangle$.
We plot the probability $P(V)$ for the ground state as a function of the magnon occupied volume $V$ (see Fig.~\ref{fig:multi_magnons}). From Fig.~\ref{fig:multi_magnons}(a), (b) and (c) it is noted that all the lines have a maximum value, and we define the corresponding value of $V$ as $V_p$. For the two-magnon system shown in Fig.~\ref{fig:multi_magnons}(a), we can see that $V_p=31$ when $g<-0.156$, in this case the two magnons tend to be separated and the effective interaction between magnons is repulsive. For $g>-0.156$, as $g$ increases, $V_p$ decreases to 2, which means the two magnons tend to condense and the effective interaction between magnons becomes attractive.
For the threshold value of $g=-0.156$, $P(V)$ is almost flat, indicating that the magnons are effectively free.
In Fig.\ref{fig:multi_magnons}(b), (c) we show the distribution of magnons with three and four magnons.  When $g=-0.8$, $V_{p}=34$ for three-magnon case and $V_{p}=40$ for four-magnon case. The large value of $V_{p}$ means that the magnons prefer to disperse.
With increase of $g$, $V_{p}$ shifts toward small value. Up to $g=0.8$, for three magnons case $V_{p}=5$ and for four magnons case $V_{p}=7$. This result indicates that the magnons tend to form a many magnon bound state.

In Fig.~{\ref{fig:multi_magnons}}(d) we plot $V_{p}$ as a function of the anisotropy $g$. It is shown that the $V_{p}$ shifts toward small value with the increase of $g$, which means the magnons have a trend to form a bound state with a strong anisotropy. Moreover, for all different magnons cases, the $V_{p}$ has a dramatic drop for certain $g$, which indicates that the formation of the bound state is quite rapid. This result has a profound insight on the magnetization jump observed in the magnetization process.
\begin{figure}[!tb] 
	\centering
	\includegraphics[width=0.9\columnwidth]{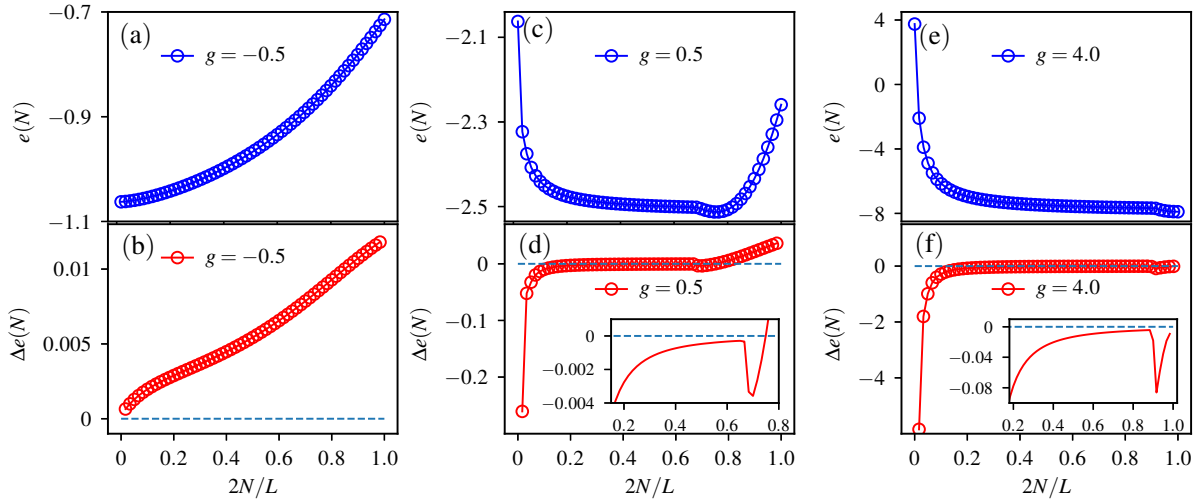}
	\caption{The energy per magnon $e(N)$ in (a) the N-MJ phase ($g=-0.5$), (c) the PF-MJ phase ($g=0.5$), and (e) the NF-MJ phase ($g=4.0$). (b), (d) and (f) are the corresponding energy difference $\Delta e(N)$ for (a), (c), and (e), respectively. For all the curves $Q=1.5$ and $L=120$.}
	\label{fig:energy_per_magnon}
\end{figure}

\subsection{The magnetization jump in the whole magnetization process}
The analysis of the effective interaction between magnons in the few magnon limit already gives a clue to the origin of the magnetization jump. Furthermore, in this subsection, we explicitly investigate the magnetization process in the presence of the external field. In this case, the arbitrary $N$-magnon state has to be considered. The energy of the $N$-magnon state subjected to the external field $h$ is
\begin{eqnarray}
E(N,h)=E(N)-h\langle S^z_{tot}\rangle,
\label{energy_in_field}
\end{eqnarray}
where the magnon number $N$ is equal to the number of the down spins, and $\langle S^z_{tot}\rangle=\sum_i \langle S^z_i\rangle$ is equal to $L/2-N$.
For simplicity, we use $E(N)$ instead of $E(N,0)$ here and hereafter.

\begin{figure}[!tb] 
	\centering
	\includegraphics[width=0.7\columnwidth]{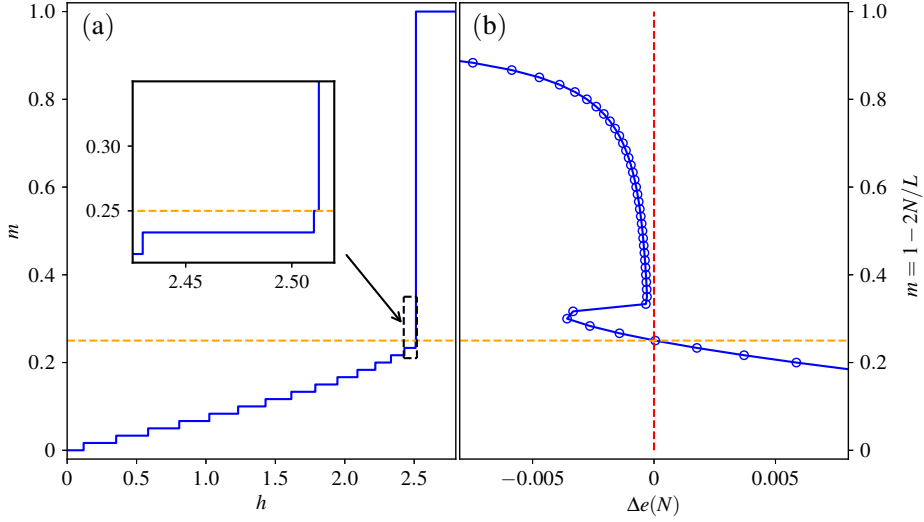}
	\caption{Comparison between (a) the magnetization curve and (b) the rotated plot for $\Delta e(N)$ as a function of $m=1-2N/L$. The $N$-magnon states with positive (negative) $\Delta e(N)$ correspond to the continuous part (sharp jump) of the magnetization curve. Here $g=0.5$, $Q=1.5$, and $L=120$.}
	\label{fig:comparison}
\end{figure}

Consider one $N$-magnon state as the ground state of the system at some external magnetic field $h$ during the magnetization process, then the ground state energy $E(N,h)$ should satisfy $E(N,h) < E(M,h)$ for any $M \neq N$. Note the fact that for the model we investigate, the magnetization density $m$ increases monotonically as $h$ increases, and there is only one jump from some critical magnetization $m_c$ to the saturated magnetization. Therefore, the condition $E(N,h) < E(M,h)$ can be rewritten as
\begin{equation}
E(N,h) < E(0,h) \label{neq01}
\end{equation}
for $M < N$, and
\begin{equation}
E(N,h)<E(N+1,h) \label{neq02}
\end{equation}
for $M > N$. Inserting Eq.~(\ref{energy_in_field}) into the conditions Eqs.~(\ref{neq01}) and (\ref{neq02}), one can easily obtain the necessary requirement of the external field $h$:
\begin{eqnarray}
h&<&-\tilde{E}(N)/N, \label{neq11} \\
h&>&E(N)-E(N+1). \label{neq12}
\end{eqnarray}
where $\tilde{E}(N)$ is defined in Eq.(\ref{eq:tilde_E}).
Combining Eqs.~(\ref{neq11}) and (\ref{neq12}), one further obtains
\begin{eqnarray}
e(N)<e(N+1),
\label{condition}
\end{eqnarray}
where $e(N) = \tilde{E}(N)/N$ is the excitation energy per magnon for the $N$-magnon state in the absence of $h$.
If the relationship in Eq.~(\ref{condition}) can not be satisfied, the $N$-magnon state can never be the ground state during the magnetization process. This is the origin of the macroscopic magnetization jump, from the perspective of the energy. More specifically, we can define the difference of the excitation energy per magnon as $\Delta e(N)=e(N+1)-e(N)$ as the determination condition of the $N$-magnon state during the magnetization process. When $\Delta e(N)>0$, the $N$-magnon state can be the ground state, and corresponds to the continuous part of the magnetization curve. Oppositely, the $N$-magnon state with $\Delta e(N)<0$ cannot be the ground state, and corresponds to the macroscopic magnetization jump.  By taking $N=2$, one can also understand the reason why the phase boundary between the N-MJ and PF-MJ phases can be determined by comparing the excitation energies in the few magnon limit.

In Fig.~\ref{fig:energy_per_magnon}, we show the excitation energy per magnon $e(N)$ and the energy difference $\Delta e(N)$ for $Q=1.5$ and several different $g$ as examples. Here $e(N)$ and $\Delta e(N)$ are numerically obtained by DMRG for each $N$-magnon states. In the N-MJ phase (e.g. $g=-0.5$), where the magnetization curve of the system is smooth and continuous (see Fig.~\ref{fig:mz}(c)), the excitation energy per magnon $e(N)$ increases monotonically as the number $N$ increases, as shown in Fig.~\ref{fig:energy_per_magnon}(a).
In this case, the energy difference $\Delta e(N)$ shown in Fig.~\ref{fig:energy_per_magnon}(b) is always positive, i.e., Eq.~(\ref{condition}) is always satisfied. We also notice that $e(N)>e(1)$ holds for all these states. It means that the energy of the $N$-magnon state is larger than $N$ free magnons. In this sense the effective interactions between magnons is always repulsive.

In the PF-MJ phase ($g=0.5$), as shown in Fig.~\ref{fig:energy_per_magnon}(c), as $N$ increases, the excitation energy per magnon $e(N)$ decreases for smaller $N$ but increases for larger $N$. As shown in Fig.~\ref{fig:energy_per_magnon}(d), there exists a region where the energy difference $\Delta e(N)<0$ , and adding a magnon to the $N$-magnon state will decreases the average energy of the magnons. This indicates the condensation of magnons, and the formation of the magnetic domain in these $N$-magnon states. These states can not be the ground state of the system in the magnetization process, and correspondingly the magnetization curve has a macroscopic jump.

Figs.~\ref{fig:energy_per_magnon}(e) and (f) show the results for the NF-MJ phase ($g=4.0$) with the magnetization jump from $m=0$ to $1$. In this phase, the excitation energy per magnon $e(N)$ decreases monotonically as the number $N$ increases, and the energy difference $\Delta e(N)$ is negative for arbitrary magnetization density.

We can further understand $\Delta e(N)$ in a more explicit way, by directly comparing the magnetization curve and the energy difference $\Delta e(N)$ as a function of $N$. Notice that magnetization $m=1-2N/L$, so Fig.~\ref{fig:comparison} (a) and (b) indeed have the same $y$-axis. As shown in Fig.~\ref{fig:comparison}(a), for a magnetization curve in the PF-MJ phase, there is a macroscopic jump from a critical $m_c$ to $m=1$. Correspondingly, the value of $\Delta e(N)$ shown in Fig.~\ref{fig:comparison}(b) has a transition from positive to negative at exactly same critical magnetization density $m_{c}$. The accordance of $m_c$ is marked by the horizontal dashed line.  Moreover, by considering the critical case of Eq.~(\ref{neq11}), we can also get the critical field $h_{\rm sat}=-e(N)$, where the critical magnon number $N$ satisfies $\Delta e(N-1)<0<\Delta e(N)$.

We can retrieve the magnetization phase diagram by plotting the critical magnetization $m_c$ in the parameter space \{$Q$, $g$\}, as shown in Fig.~\ref{fig:phase_diag}(b). When the magnetization curve is smooth and continuous, $m_c$ should be $1$ in the thermodynamic limit, indicating there is no magnetization jump. However, for the finite size system, we have $m_c=1-2/L$ because of a microscopic quantized jump. Nevertheless, the N-MJ phase denoted by the darkest blue is distinct in Fig.~\ref{fig:phase_diag}(b). For a fixed $g>(-4+\sqrt{7})/3$, the magnetization jump appears as $Q$ increases to the critical value, and $m_{c}$ decreases with the increasing of $Q$. Finally, when $g$ and $Q$ are both sufficiently large, the system is in the NF-MJ phase with $m_c=0$. All these phases and the corresponding phase boundaries are explicit and clear.

\subsection{Understanding the direct magnetization jump in large anisotropy limit }
In the macroscopic viewpoint, the direct magnetization jump can be understood in an analytical and intuitive way in the large $g$ limit. When the anisotropy is large enough, the system enters into an NF-MJ phase, as shown in Fig. \ref{fig:phase_diag}. In this limit, being divided by $g^{2}$ on both sides, the Hamiltonian described by Eq.~(\ref{ham_jq}) reads(details in supplementary material)
\begin{equation}
H/g^2  = -Q\sum_{i}S_{i}^{z}S_{i+1}^{z}S_{i+2}^{z}S_{i+3}^{z} + O\left(1/g\right) + O\left(1/g^2\right)-h^{\prime}\sum_{i}S_{i}^{z},
\end{equation}
where $h^{\prime}=h/g^{2}$.
By neglecting the $O(1/g)$ and $O(1/g^{2})$ terms, we have an effective Hamiltonian in the large $g$ limit
\begin{equation}\label{eq:limit_ham}
\mathcal{H}_{g\rightarrow\infty}=-Q\sum_{i}S_{i}^{z}S_{i+1}^{z}S_{i+2}^{z}S_{i+3}^{z}-h^{\prime}\sum_{i}S_{i}^{z}.
\end{equation}
Equation~(\ref{eq:limit_ham}) describes a classical Hamiltonian without quantum fluctuation, then we can easily get the ground state energy and the spin configuration of the system. The unit element of this Hamiltonian is a bond with 4 sites, and the total energy of the system is the summation of all the bonds. A bond contributes negative energy $-E_b$ when the numbers of both up and down spins are even, where $E_b = Q/16$ as the bond energy. Oppositely, when the numbers of both up and down spins are odd, a bond has positive energy $+E_b$. We have listed all possible spin arrangements of a single bond in Table~\ref{table:bond}.

\begin{table}
\centering
\begin{tabular}{c|c}
\hline
Bond energy & Possible spin configurations \tabularnewline
\hline
\hline
$-E_{b}$ & $\uparrow\uparrow\uparrow\uparrow$ \tabularnewline
\cline{2-2}
 & $\downarrow\downarrow\downarrow\downarrow$\tabularnewline
\cline{2-2}
 & $\uparrow\downarrow\uparrow\downarrow$, $\downarrow\uparrow\downarrow\uparrow$\tabularnewline
\cline{2-2}
 & $\uparrow\uparrow\downarrow\downarrow$, $\uparrow\downarrow\downarrow\uparrow$, $\downarrow\downarrow\uparrow\uparrow$, $\downarrow\uparrow\uparrow\downarrow$\tabularnewline
\hline
$+E_{b}$ & $\uparrow\uparrow\uparrow\downarrow$, $\uparrow\uparrow\downarrow\uparrow$, $\uparrow\downarrow\uparrow\uparrow$, $\downarrow\uparrow\uparrow\uparrow$\tabularnewline
\cline{2-2}
 & $\downarrow\downarrow\downarrow\uparrow$, $\downarrow\downarrow\uparrow\downarrow$, $\downarrow\uparrow\downarrow\downarrow$, $\uparrow\downarrow\downarrow\downarrow$\tabularnewline
\hline
\end{tabular}
\caption{The energy and possible spin configurations for a single bond of the effective Hamiltonian described by Eq.~(\ref{eq:limit_ham}). }
\label{table:bond}
\end{table}

Without loss of generality, we consider the system with even sites $L$ with external magnetic field $h^{\prime}=0$. The ground state of the system has magnetization $m=0$ or $m=\pm1$, with ground state energy $E_{g\rightarrow\infty}=-L E_b$, since there are $L$ bonds under PBCs. For $m=0$, the spin configuration can be a 2-fold degenerated spin pattern $\left|\cdots\uparrow\downarrow\uparrow\downarrow\cdots\right\rangle$, or a 4-fold degenerated spin pattern $\left|\cdots\uparrow\uparrow\downarrow\downarrow\cdots\right\rangle$. For $m=1(-1)$, the spin configuration can be $\left|\cdots\uparrow\uparrow\uparrow\uparrow\cdots\right\rangle$ ($\left|\cdots\downarrow\downarrow\downarrow\downarrow\cdots\right\rangle$). Introducing infinitely small quantum fluctuations, the ground state has $m=0$ when $h^{\prime}=0$, and will be fully-polarized under a small magnetic field. Thus, the direct jump is  the only choice for the magnetization process.

Furthermore, we consider the \textit{jumped-over} spin states with magnetization $0<m<1$. To minimize the energy, the spin pattern has to be separated into two regions: i) a magnon-full region with $m=0$ and ii) a fully-polarized domain region with $m=1$. Therefore in this case, all the bonds within the same region have negative energy, and only the bonds across the two regions can contribute positive energy. For example, the  spin structure can be  $\left|\cdots\downarrow\downarrow\uparrow\uparrow + \uparrow\uparrow\uparrow\uparrow \cdots\right\rangle$, and only the bond $\left|\cdots\downarrow\uparrow\uparrow + \uparrow \cdots\right\rangle$ that connects the two separated parts of the system contributes positive energy $+E_b$.
Therefore, in the large $g$ limit, for all the \textit{jumped-over} states with magnetization $0<m<1$, its ground state has magnetic domains. For this special model, we can also conclude that all the states with magnetic domain can not be the ground state of the system. In other words, considering the magnetization process, all the states with magnetic domain will be jumped over during the magnetization process. We expect this point is not only valid for the large $g$ limit, but also be crucial for a general value of $g$.

\subsection{Correlation functions}

According to the previous subsections, we found that the states with magnetization domain structure are jumped over during the magnetization process. In this subsection, we are interested in the difference between the structures of the \textit{jumped-over} states and experienced states.
To unveil the physics of the magnetization jump beyond the energy perspective, we investigate the spin-spin correlation function:
\begin{eqnarray}
C_{S}(r)=\left<S^z_0 S^z_r\right>-\left<S^z_0\right>\left<S^z_r\right>,
\end{eqnarray}
where $r$ is real space coordinate.

\begin{figure}[!tb] 
	\centering
	\includegraphics[width=0.9\columnwidth]{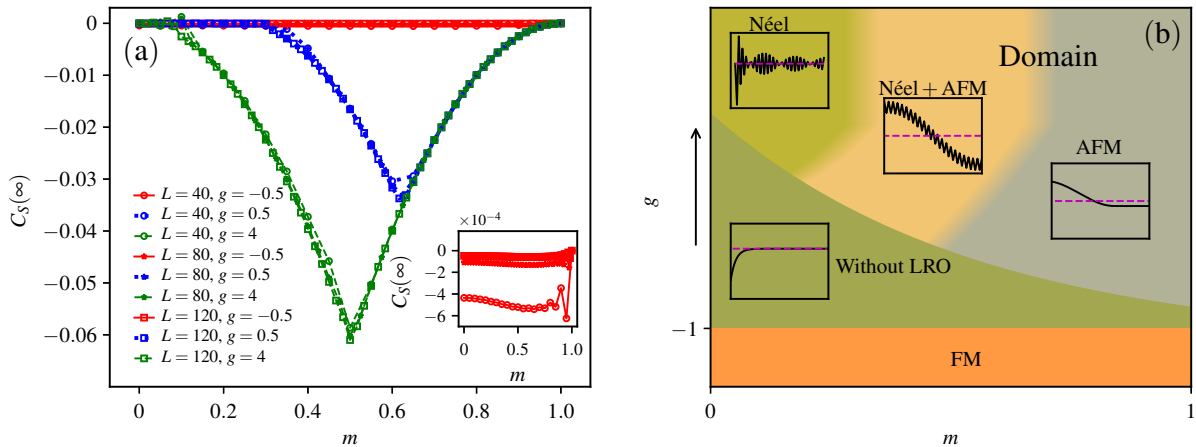}
	\caption{ (a) Spin-spin correlation function in the long-range limit as a function of $m=1-2N/L$ for $Q=1.5$, different anisotropy $g$, and different system sizes. The inset is a zoom-in for $g=-0.5$. (b) The schematic phase diagram for a fixed nonzero $Q$ in the absence of external field $h$. In each inset, the black solid-line represents the schematic spin-spin correlation function $C_{S}\left(r\right)$ (details in supplementary material), and the magenta dashed-line denotes $C_{S}(r)=0$.}
	\label{fig:long_range_order}
\end{figure}

We plot the long-range correlation function $C_S(\infty)$ as a function of magnetization density $m=1-2N/L$ in Fig.~\ref{fig:long_range_order}(a). Here we define $C_S(\infty)=[C_S(L/2)+C_S(L/2-1)]/2$ to remove the strong oscillations when $g$ is very large. For the N-MJ phase without magnetization jump, $C_S(\infty)$ is very small for all the magnetization densities, and its amplitude decreases as $L$ increases (see inset). Therefore, in the thermodynamic limit $C_S(\infty)$ is zero, and there is no LRO in this phase.

In the PF-MJ phase ($g=0.5$), $C_S(\infty)$ approaches 0 for small magnetization densities, where the magnetization curve is continuous. For these \textit{jumped-over} states at larger $m$, $C_S(\infty)$ is nonzero and show convergence for different system sizes. Therefore in the thermodynamic limit, the \textit{jumped-over} state has AFM-LRO because of the formation of magnetic domain.

In the NF-MJ phase ($g=4.0$), $C_S(\infty)$ is nonzero for larger magnetization density. Specially, for $m$ between 0.7 and 1, $C_S(\infty)$ is the same as in the PF-MJ phase independent of the system size, as these $N$-magnon states share the same domain structure. However, different with the PF-MJ phase, the spin-spin correlation function has large oscillations in the long-range limit for those states with magnetization densities from $m = 0.1$ to $0.3$.
The states near $m=0$ have nearly zero $C_S(\infty)$, there is no AFM-LRO or domain, but the spin-spin correlation function has long-range N\'eel oscillations (details in supplementary material), as large $g$ drives the system to the classical limit.

\section{Conclusion}
\label{sec_5}

In this work, we systematically investigate the adiabatic magnetization properties of the 1D anisotropic $J-Q_{2}$ model at zero temperature by numerically using the DMRG method.
We have found that the anisotropy $g$ plays a crucial role in the magnetization process. The characteristics of the magnetization behavior can be summarized by a magnetization phase diagram consisting of four phases: the FM phase, the N-MJ phase without magnetization jump, the PF-MJ phase with a partially- to fully-polarized magnetization jump, and specially the NF-MJ phase with a direct magnetization jump from non- to fully-polarized state, which does not exist in the isotropic $J-Q_{2}$ model.

We further study the origin of the magnetization jump. In the few magnon limit, we analyse the system with up to four magnons and get the clue that the attractive interaction between magnons may effects the formation of magnetization jump.
For the $N$-magnon state, we point out that the origin of the magnetization jump is the condensation of magnons from the energy consideration. For the direct magnetization jump which is absent in the isotropic system, the analysis in the limit of infinite large anisotropy shows that the magnetization domain plays an important role in the magnetization jump. By explicitly investigating the spin-spin correlation function, we confirm that the spins condense and form the magnetic domain in those \textit{jumped-over} states. A schematic phase diagram is shown in Fig.~\ref{fig:long_range_order}(b) for a fixed non-zero pair coupling: i) If the magnetization curve is continuous, the corresponding ground states of the system cannot have any long-range order; ii) The state with long-range orders (e.g. antiferromagnetic or N\'eel long-range orders, or their mixing) cannot be the ground state of the system during the magnetization process, and therefore the magnetization jump arises. This reminds us the fact that the 1D spin-1/2 chain cannot have a stable long-range ordered ground state \cite{Landau1958} with continuous symmetry breaking due to the strong quantum fluctuations. Therefore, the conclusion obtained here is not only valid to the $J-Q_{2}$ model we study, but also should be a general conjecture for a wide range of 1D spin models and materials.

\bibliography{sample}

\section*{Acknowledgments}

The authors acknowledge useful discussions with D-X. Yao, H. Shao, W-A. Guo, L. Wang and M. Liu.
H-G. Luo acknowledges the support from NSFC (Grants No. 11325417, 11674139) and PCSIRT (Grant No. IRT-16R35).
C. Cheng acknowledges support from NSAF U1530401 and computational resource from the Beijing Computational Science Research Center.

\section*{Author contributions statement}

B.M. designed the work under the guide of H.L. and carried out the calculations. H.L., B.M. and C.C. analysed the data and wrote the manuscript. F.C. wrote part of DMRG code. H.L. supervised the work. All authors reviewed the manuscript.

\section*{Additional information}
\subsection*{Supplementary information}
accompanies this paper at http://www.nature.com/srep

\subsection*{Competing financial interests:} The authors declare no competing financial interests.
\subsection*{Data availability statement:}
The datasets generated during and/or analysed during the current study are available from the corresponding author on reasonable request.

\newpage
\section{Supplementary material}
\subsection{Hamiltonian in the few magnon limit}
Note that in this study the macroscopic magnetization jump always happens from a finite magnetization to the saturated magnetization state, which is the ferromagnetic state. Therefore, to understand the magnetization jump in the simplest way, it is natural to consider the one- and two-magnon states on a ferromagnetic background. The ferromagnetic state, which is a zero-magnon state, is denoted as $|0\rangle=|\uparrow\uparrow\uparrow\cdots\rangle$. In the Following, we examine the ground state for the system in $N$-magnon sectors in the few magnon limit (up to $N=2$), neglecting the contribution of the external field. For simplicity, we denote the Hamiltonian as
\begin{equation}\label{ham_jq_1}
H=H_J + H_Q,
\end{equation}
where $H_J=-J\sum_{i}P_{i,i+1}$ and $H_Q=-Q\sum_{i}P_{i,i+1}P_{i+2,i+3}$. For the system described by the Hamiltonian in Eq.~(\ref{ham_jq_1}), with size $L$ and periodic boundary conditions (PBC), the energy of this zero-magnon state is
\begin{equation}
E(0)=-J\frac{\left(1-g\right)}{4}L-Q\frac{\left(1-g\right)^{2}}{16}L.
\end{equation}

The one-magnon excited state with momentum $k$ can be defined as
\begin{equation}
\left|k\right\rangle=\frac{1}{\sqrt{L}}\sum_{l=1}^{L}e^{ikl}S_{l}^{-}\left|0\right\rangle.
\end{equation}
By acting the Hamiltonian on the state, we get
\begin{equation}
H_{J}\left|k\right\rangle   =  J \cos k\left|k\right\rangle -\frac{J\left(1-g\right)L}{4}\left|k\right\rangle -Jg\left|k\right\rangle
\end{equation}
for $J$ term, and
\begin{equation}
H_{Q}\left|k\right\rangle =\left[\frac{Q\left(1-g\right)}{2} \cos k-\frac{Q\left(1-g\right)^{2}L}{16}+\frac{Q\left(g^{2}-g\right)}{2}\right]\left|k\right\rangle
\end{equation}
for $Q$ term. Notice that $H$ is diagonal in $|k\rangle$ basis, we can easily obtain the energy dispersion of the system with one-magnon
\begin{equation}
E_{k}(1)   =  \left[J+\frac{Q}{2}\left(1-g\right)\right]{ \cos}\left(k\right)-Jg+\frac{Q\left(g^{2}-g\right)}{2}-\frac{J\left(1-g\right)L}{4}-\frac{Q\left(1-g\right)^{2}L}{16}.
\end{equation}
As in the main text, define $N$ magnon excitation energy as $\tilde{E}(N)=E(N)-E(0)$. The one-magnon excitation energy is
\begin{equation}
\tilde{E}(1) \!=\!
\begin{cases}
-J\left(1+g\right)\!-\!\frac{Q\left(1-g^{2}\right)}{2} &  \text{if $g<\frac{2J}{Q}+1$,$k=\pi$} \\
J\left(1-g\right)+\frac{Q\left(1-g\right)^{2}}{2}  &  \text{if $g>\frac{2J}{Q}+1$,$k=0$.}
\end{cases}
\end{equation}
Here $\tilde{E}(1)$ can also be considered as the energy of a free magnon.

For two-magnon state, we choose the basis with a total momentum $k$ and a relative distance $d$ defined as
\begin{equation}
\left|d,k\right\rangle =\frac{1}{\sqrt{L}}\sum_{l}e^{ik\left(2l+d\right)/2}S_{l}^{-}S_{l+d}^{-}\left|0\right\rangle.
\end{equation}

Acting the Hamiltonian on the basis set, for the $J$ term, we obtain:
\begin{equation}
H_{J}\left|1,k\right\rangle   =  \left(\frac{g-1}{4}JL-Jg\right)\left|1,k\right\rangle +J\cos \frac{k}{2}\left|2,k\right\rangle,
\end{equation}
\begin{equation}
H_{J}\left|d>1,k\right\rangle   = J\cos \frac{k}{2}\left(\left|d+1,k\right\rangle +\left|d-1,k\right\rangle \right)+\left(\frac{g-1}{4}JL-2Jg\right)\left|d,k\right\rangle.
\end{equation}

For the $Q$ term, there are
\begin{equation}
H_{Q}\left|1,k\right\rangle   =  \left[-\frac{\left(1-g\right)^{2}QL}{16}+\frac{\left(g^{2}-2g\right)Q}{4}\right]\left|1,k\right\rangle +\frac{Q}{2}\cos \frac{k}{2}\left|2,k\right\rangle -\frac{Q}{4}\left|3,k\right\rangle,
\end{equation}
\begin{equation}
H_{Q}\left|2,k\right\rangle   =  \frac{Q}{2}\cos \frac{k}{2}\left(\left|3,k\right\rangle +\left|1,k\right\rangle \right) +\left[-\frac{\left(1-g\right)^{2}QL}{16}+\frac{\left(g^{2}-2g-\cos k\right)Q}{2}\right]\left|2,k\right\rangle,
\end{equation}
\begin{equation}
H_{Q}\left|3,k\right\rangle   =  -\frac{Q}{4}\left|1,k\right\rangle +\frac{Q}{2}\cos \frac{k}{2}\left|2,k\right\rangle
+\left[-\frac{\left(1-g\right)^{2}QL}{16}+Q\left(\frac{3}{4}g^{2}-g\right)\right]\left|3,k\right\rangle  +\frac{Q\left(1-g\right)}{2}\cos \frac{k}{2}\left|4,k\right\rangle,
\end{equation}
\begin{equation}
H_{Q}\left|d>3,k\right\rangle   =  \frac{Q\left(1-g\right)}{2}\cos \frac{k}{2}\left(\left|d+1,k\right\rangle +\left|d-1,k\right\rangle \right) +\left[-\frac{\left(1-g\right)^{2}QL}{16}+Q\left(g^{2}-g\right)\right]\left|d,k\right\rangle.
\end{equation}

\begin{flushleft}
	\begin{figure}[!tb] 
		\centering
		\includegraphics[width=0.6\columnwidth]{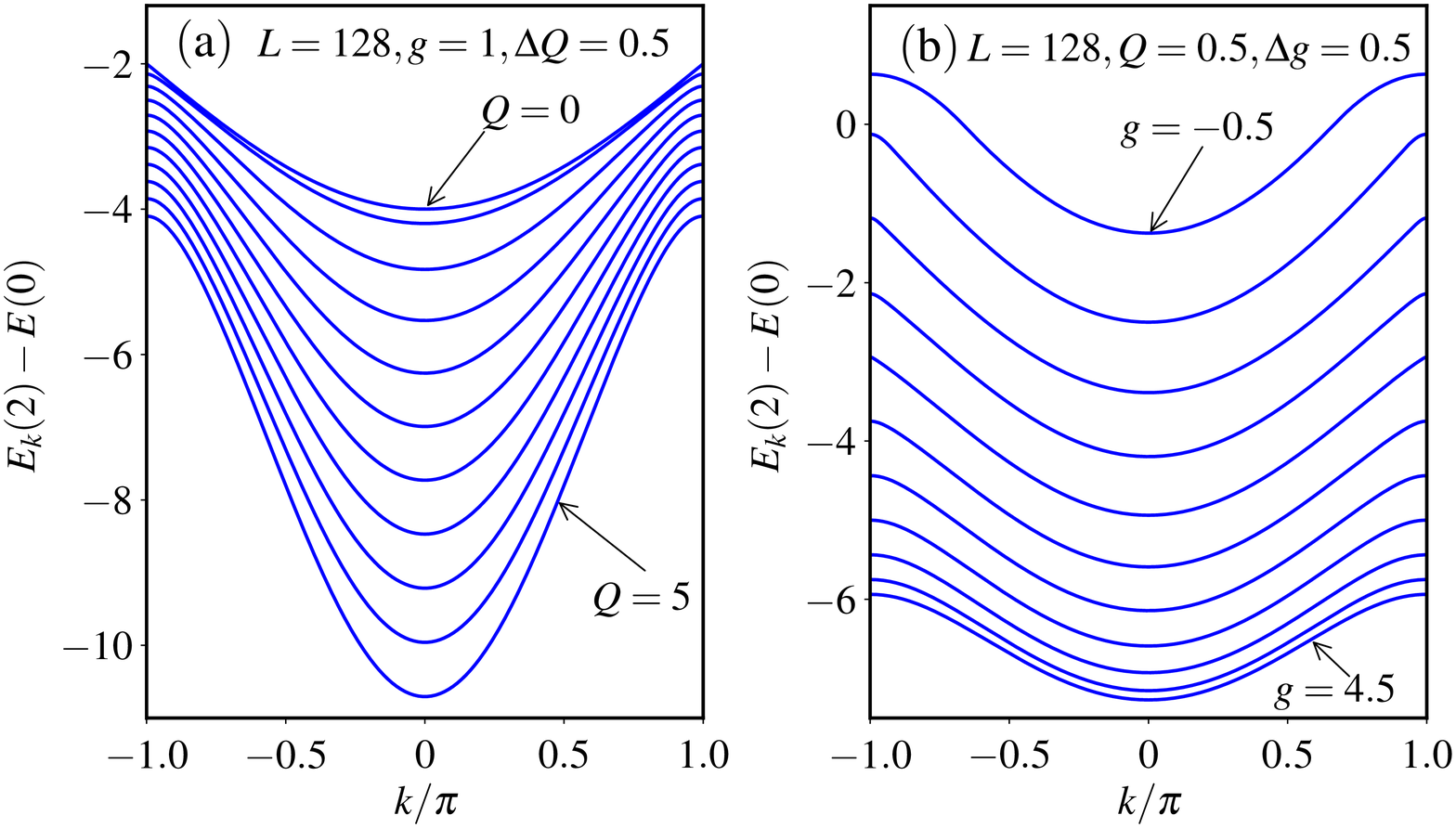}
		\caption{ (Color online) The dispersion of the two-magnon exited states of (a) $g=1$ for different $Q$, and (b) $Q=0.5$ for different $g$. Here system size $L=128$. }
		\label{fig:E2k}
	\end{figure}
\end{flushleft}

Then the ground state energy $E_2(k)$ of the two-magnon state with momentum $k$ can be obtained by numerically diagonalizing the $(L-1)\times(L-1)$ Hamiltonian matrix in the basis set $\left|d,k\right\rangle$. In Fig.~\ref{fig:E2k} we show the dispersion $E_k(2)-E(0)$ of the two-magnon exited states for several different parameters. As one can see, for these examples the two-magnon ground state always has $k=0$. Actually, we have carefully checked the dispersion for all the parameters we concern in this work, and the minimum $E_k(2)-E(0)$ for each point in the parameter space always has zero momentum. Therefore, the Hamiltonian matrix $\tilde{H}(2)=H_J+H_Q-E(0)$ in the two-magnon basis can be simplified as
\begin{align}
\tilde{H}(2) =	\begin{pmatrix}
Q\frac{g^{2}-2g}{4}-Jg & \frac{Q}{2}+J & -\frac{Q}{4} &\\
\frac{Q}{2}+J & Q\frac{g^{2}-2g-1}{2}-2Jg & \frac{Q}{2}+J &\\
-\frac{Q}{4} & \frac{Q}{2}+J & Q\frac{3g^{2}-4g}{4}-2Jg & \frac{Q\left(1-g\right)}{2}+J &\\
&  & \frac{Q\left(1-g\right)}{2}+J & Q\left(g^{2}-g\right)-2Jg & \\
&  &  &  &\ddots\\
\end{pmatrix}.
\label{ham_e2}
\end{align}

\begin{flushleft}
	\begin{figure}[!tb] 
		\centering
		\includegraphics[width=0.45\columnwidth]{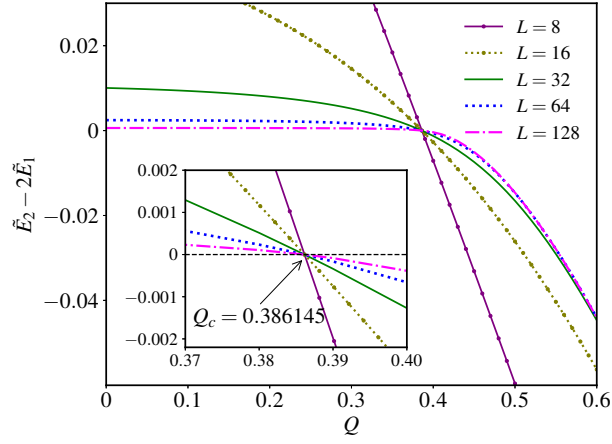}
		\caption{ (Color online) $\tilde{E}(2)-2\tilde{E}(1)$  as a function of $Q$ for $g=0.5$ and different $L$. The black dashed line in inset is for $\tilde{E}(2)-2\tilde{E}(1)=0$.}
		\label{fig:e1e2_finite_size}
	\end{figure}
\end{flushleft}

In order to see the finite size effect in the few magnon limit, we plot $\tilde{E}(2)-2\tilde{E}(1)$ for $g=0.5$ as a function of $Q$ for different system sizes, as shown in Fig.~\ref{fig:e1e2_finite_size}. All these curves have a precise cross at $\tilde{E}(2)-2\tilde{E}(1)=0$ and a critical $Q_c{(g=0.5)}=0.386145$ even for $L=8$, which is the minimum system size to include all the information of the effective Hamiltonian described by Eq.~(\ref{ham_e2}). Therefore, the finite size effect in the few magnon limit is negligible.

\subsection{The asymptotic behavior}
From Fig.3(c) in the main text we can see that the phase boundary between the N-MJ and PF-MJ phases can be exactly determined by comparing the energy of one- and two-magnon excitations. The phase boundary obtained in the few magnon limit perfectly agrees with the numerical results by DMRG.

We also notice the asymptotic behavior of this curve when the pair coupling $Q$ is extremely large. In the limit of $Q\rightarrow\infty$, the one-magnon excitation energy is
\begin{eqnarray}\label{eq:e1_q}
\centering
\tilde{E}(1)/Q =
\begin{cases}
-\frac{1-g^{2}}{2} &  \text{if $g<1$,} \\
\frac{\left(1-g\right)^{2}}{2}  &  \text{if $g>1$.}
\end{cases}
\end{eqnarray}
here we have ignored the infinite small terms proportional to $1/Q$. Similarly, ignoring the $O(1/Q)$ terms, the two-magnon excitation is described by the matrix
\begin{eqnarray}
\centering
\label{H2overQ}
\tilde{H}(2)/Q	=
&&\left(\begin{array}{cccccccc}
\frac{g^{2}-2g}{4} & \frac{1}{2} & -\frac{1}{4}\\
\frac{1}{2} & \frac{g^{2}-2g-1}{2} & \frac{1}{2}\\
-\frac{1}{4} & \frac{1}{2} & \frac{3g^{2}-4g}{4} & \frac{1-g}{2}\\
&  & \frac{1-g}{2} & g^{2}-g & \\
& & & & \ddots
\end{array}\right).
\end{eqnarray}

We can numerically obtain the critical anisotropy $g_c$ satisfying $\tilde{E}(2)-\tilde{E}(1)=0$.  For the ground state wave function ($\left|G\right\rangle_2 =\sum_{d=1}^{L-1} \alpha_{d}\left|d\right\rangle$) at the critical point, we find that i)$\alpha_{d}^2$ is a constant number, ii) $\alpha_{d+1} = -\alpha_d$,  for $d$ and $d+1$ in range $[3, L-3]$. Thus, the critical wave function can be assumed as
\begin{equation}
\left|G\right\rangle_2 =\frac{1}{\eta}\left[ a\left|1\right\rangle
+ b\left|2\right\rangle + c\left|3\right\rangle + \sum_{d=4}^{L-4}\left(-1\right)^{d}\left|d\right\rangle
+ c\left|L-3\right\rangle
+b\left|L-2\right\rangle
+a\left|L-1\right\rangle\right],
\end{equation}
where $\left|d\right\rangle\equiv\left|d,k=0\right\rangle$, $\eta$ is the normalization coefficient.

Applying the Hamiltonian in Eq.~(\ref{H2overQ}) to the wavefunction $|G\rangle_2$,we can get a set of equations.
By solving them, the critical $g$ in the $Q\rightarrow\infty$ limit can be obtained as $g_c(Q\rightarrow\infty)=\left( -4+ \sqrt{7} \right)/3$.
For any anisotropy $g$ below this value, the magnetization curve of the system is always smooth and continuous.

\subsection{The effective Hamiltonian in large anisotropy limit}
Divided by $g^{2}$ on both sides, the Hamiltonian in this study can be written as
\begin{equation}
\frac{H}{g^{2}}  = -Q\sum_{i}S_{i}^{z}S_{i+1}^{z}S_{i+2}^{z}S_{i+3}^{z}+O\left(\frac{1}{g}\right)+O\left(\frac{1}{g^{2}}\right)-h^{\prime}\sum_{i}S_{i}^{z},
\end{equation}
where
\begin{equation}
h^{\prime}=\frac{h}{g^{2}},
\end{equation}
\begin{equation}
\begin{split}
O\left(\frac{1}{g}\right)  = &\frac{1}{g}\Bigg\{ J\sum_{i}S_{i}^{z}S_{i+1}^{z} \\
& +Q\sum_{i}S_{i}^{z}S_{i+1}^{z}\left[\frac{1}{4}-\frac{1}{2}\left(S_{i+2}^{+}S_{i+3}^{-}+S_{i+2}^{-}S_{i+3}^{+}\right)\right]\\
& +Q\sum_{i} \left[\frac{1}{4}-\frac{1}{2}\left(S_{i}^{+}S_{i+1}^{-}+S_{i}^{-}S_{i+1}^{+}\right)\right]S_{i+2}^{z}S_{i+3}^{z}\Bigg\},\\
\end{split}
\end{equation}
\begin{equation}
\begin{split}
O\left(\frac{1}{g^{2}}\right)=&\frac{1}{g^{2}}\Bigg\{-J\sum_{i}\left[\frac{1}{4}-\frac{1}{2}\left(S_{i}^{+}S_{i+1}^{-}+S_{i}^{-}S_{i+1}^{+}\right)\right] \\
& -Q\sum_{i}\left[\frac{1}{4}-\frac{1}{2}\left(S_{i}^{+}S_{i+1}^{-}+S_{i}^{-}S_{i+1}^{+}\right)\right]\left[\frac{1}{4}-\frac{1}{2}\left(S_{i+2}^{+}S_{i+3}^{-}+S_{i+2}^{-}S_{i+3}^{+}\right)\right]\Bigg\}.\\
\end{split}
\end{equation}
In the limit of $g\rightarrow\infty$, $Q$ is finite, the $O(1/g)$ and $O(1/g^{2})$ terms can be neglected . Thus the effective Hamiltonian in the large anisotropy limit is
\begin{equation}
\mathcal{H}_{g\rightarrow\infty}=-Q\sum_{i}S_{i}^{z}S_{i+1}^{z}S_{i+2}^{z}S_{i+3}^{z}-h^{\prime}\sum_{i}S_{i}^{z}.
\end{equation}

\subsection{The correlation function}
\begin{flushleft}
	\begin{figure}[!tb] 
		\centering
		\includegraphics[width=0.9\columnwidth]{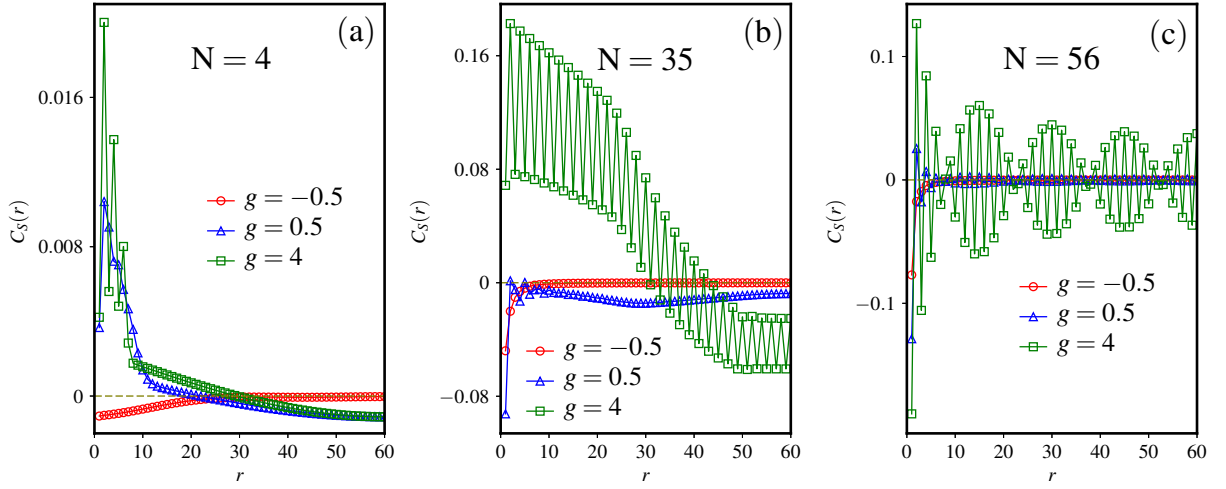}
		\caption{ Spin-spin correlation function $C_{S}(r)$ for different $g$ and $N$. Here $Q=1.5$ and $L=120$. }
		\label{fig:correlation_1}
	\end{figure}
\end{flushleft}
In this section we discuss the different behaviors of the spin-spin correlation function
in the experienced sectors and the \textit{jumped-over} sectors. Fig.~\ref{fig:correlation_1} displays the correlation function $C_{S}(r)$ with different parameters as examples. In the N-MJ phase ($g=-0.5$), independent of the magnon number $N$, the spin-spin correlation $C_S(r)$ is negative for all the distance $r>0$, and rapidly decays to $0$ as $r$ increases. In this case, a spin has anti-ferromagnetic correlation with its environment, and is screened due to the strong quantum fluctuation. The states in the experienced sector have no long-range order (LRO).

In the PF-MJ phase ($g=0.5$), the magnetization curve is continuous at some magnetization density, and then has a sharp jump. The blue triangle-line shown in Fig.~\ref{fig:correlation_1}(a) for $N=4$ is the correlation function of an \textit{jumped-over} state. In this state, $C_{S}(r)$ is positive when the distance $r$ is small but negative when the spins are far apart from each other, which is the typical behavior of the system having two ferromagnetic domains. We also notice $C_{S}(r)$ has a finite value even when $r=L/2$, which is the largest distance possible for the finite system with system size $L$. In fact, $C_{S}(r)$ seems to converge when $r$ is large enough. This indicates the anti-ferromagnetic (AFM) long-range order of the \textit{jumped-over} states. The AFM-LRO still can be observed when $N=35$, as this state is also jumped over in the magnetization process, as the blue triangle-line shown in Fig.~\ref{fig:correlation_1}(b). Further increasing the magnon number $N$, the correlation function $C_{S}(r)$ of the $56$-magnon state decays rapidly to zero, similar to the situation in the N-MJ phase. The AFM-LRO disappears, and the magnetization curve in this region is continuous.

The correlation functions are more complicated in the NF-MJ phase ($g=4.0$). Since the anisotropy $g$ is sufficient large, the diagonal term of the Hamiltonian dominates and the quantum effect has been partially depressed. We can observe the strong fluctuation of $C_S(r)$ at very large distance, especially when the magnon number $N$ is large. Nevertheless, when $N$ is small ($N=4$ and $35$), besides the fluctuations, the long-range nature of domain wall is still true at large distance. For a large $N=56$, which is near the zero magnetization, the correlation function shown in Fig.~\ref{fig:correlation_1}(c) exhibits a classical N\'eel order.
\end{document}